\documentclass[twocolumn,floatfix,superscriptaddress,a4paper,showpacs,showkeys,nofootinbib,reprint,prc,noeprint]{revtex4-1}
\usepackage{epsfig}
\usepackage{latexsym}
\usepackage[colorlinks=true,linktocpage=true,linkcolor=blue,citecolor=blue,allcolors=blue]{hyperref}
\usepackage{url}
\usepackage[utf8]{inputenc}
\usepackage{enumerate}
\usepackage{color}
\usepackage{xcolor}

\usepackage{xfrac}

\usepackage{amsmath}
\usepackage{amssymb}

\usepackage[english]{babel}
\usepackage{hyperref}
\usepackage{url}
\usepackage{needspace}
\usepackage{float}
\hypersetup{
   colorlinks=true, %set true if you want colored links
  linktoc=all,     %set to all if you want both sections and subsections linked
  linkcolor=blue,  %choose some color if you want links to stand out
}

\begin{document}

 \title{Nuclei and hypernuclei production in pion induced reactions around threshold energies}

\author{Apiwit Kittiratpattana}
\affiliation{Institut f\"{u}r Theoretische Physik, Goethe Universit\"{a}t Frankfurt, Max-von-Laue-Str. 1, D-60438 Frankfurt am Main, Germany}
\affiliation{Center of Excellence in High Energy Physics and Astrophysics, School of Physics, Suranaree University of Technology, University Avenue 111, Nakhon Ratchasima 30000, Thailand}

\author{Tom Reichert}
\affiliation{Institut f\"{u}r Theoretische Physik, Goethe Universit\"{a}t Frankfurt, Max-von-Laue-Str. 1, D-60438 Frankfurt am Main, Germany}
\affiliation{Helmholtz Research Academy Hesse for FAIR (HFHF), GSI Helmholtzzentrum f\"ur Schwerionenforschung GmbH, Campus Frankfurt, Max-von-Laue-Str. 12, 60438 Frankfurt am Main, Germany}

\author{Nihal Buyukcizmeci}
\affiliation{Department of Physics, Selcuk University, 42079 Kamp\"us, Konya, T\"urkiye}

\author{Alexander Botvina}
\affiliation{Institut f\"{u}r Theoretische Physik, Goethe Universit\"{a}t Frankfurt, Max-von-Laue-Str. 1, D-60438 Frankfurt am Main, Germany}
\affiliation{Helmholtz Research Academy Hesse for FAIR (HFHF), GSI Helmholtzzentrum f\"ur Schwerionenforschung GmbH, Campus Frankfurt, Max-von-Laue-Str. 12, 60438 Frankfurt am Main, Germany}

\author{Ayut Limphirat}
\affiliation{Center of Excellence in High Energy Physics and Astrophysics, School of Physics, Suranaree University of Technology, University Avenue 111, Nakhon Ratchasima 30000, Thailand}

\author{Christoph Herold}
\affiliation{Center of Excellence in High Energy Physics and Astrophysics, School of Physics, Suranaree University of Technology, University Avenue 111, Nakhon Ratchasima 30000, Thailand}

\author{Jan Steinheimer}
\affiliation{Frankfurt Institute for Advanced Studies, Ruth-Moufang-Str. 1, D-60438 Frankfurt am Main, Germany}

\author{Marcus Bleicher}
\affiliation{Institut f\"{u}r Theoretische Physik, Goethe Universit\"{a}t Frankfurt, Max-von-Laue-Str. 1, D-60438 Frankfurt am Main, Germany}
%%\affiliation{
%%GSI Helmholtzzentrum f\"ur Schwerionenforschung GmbH, Planckstr. 1, D-64291 Darmstadt, Germany}
\affiliation{Helmholtz Research Academy Hesse for FAIR (HFHF), GSI Helmholtzzentrum f\"ur Schwerionenforschung GmbH, Campus Frankfurt, Max-von-Laue-Str. 12, 60438 Frankfurt am Main, Germany}

\date{\today}

\begin{abstract}
The Ultra-relativistic Quantum Molecular Dynamics model is employed to simulate $\pi^-+\mathrm{C}$ and $\pi^-+\mathrm{W}$ collisions at p$_\mathrm{lab}=1.7$ GeV motivated by the recent HADES results. By comparing the proton and $\Lambda$ transverse momentum spectra, it was observed that the data and transport model calculation show a good agreement, if cluster formation is included to obtain the free proton spectra. Predictions of light cluster ($d$, $t$, $^3$He, $^4$He, as well as ${}^{3}_\Lambda$H and $\Xi$N) multiplicities and spectra are made using a coalescence mechanism. The resulting multiplicities suggest that the pion beam experiment can produce a substantial amount of ${}^{3}_\Lambda$H, especially in $\pi^-+\mathrm{W}$ collisions due to the stopping of the $\Lambda$ inside the large tungsten nucleus. The findings are supplemented by a statistical multi-fragmentation analysis suggesting that even larger hyperfragments are produced copiously. It is suggested that even double strange hypernuclei are in reach and might be studied in more detail using a slightly higher pion beam momentum.
\end{abstract}

\maketitle

\section{Introduction}
The production of hypernuclei and non-strange clusters is currently under study at a large variety of heavy ion accelerator facilities. Heavy ion experiments at RHIC and LHC have recently published detailed studies on the production of hypernuclei with a special focus on light hypernuclei, e.g. the hypertriton (${}^{3}_\Lambda$H) \cite{STAR:2010gyg,Rappold:2013fic,Rappold:2014jqa,ALICE:2015oer,STAR:2017gxa,STAR:2019wjm,ALICE:2019vlx,STAR:2021orx}. It was suggested \cite{Sun:2018mqq} that the specific structure of the hypertriton, consisting of a deuteron core with a weakly bound $\Lambda$, would result in a specific system size dependence of the hypertriton production yield. Generally, however, the data situation for hypernuclei production is rather limited especially for small systems. While at high energies and large systems new experimental data on light hypernuclei and non-strange cluster production is available, the situation for smaller systems and lower collision energies is different. Here, the HADES experiment \cite{HADES:2009aat} can close a gap with its recently measured data using a pion beam with a momentum of 1.7 GeV on a carbon (C) and tungsten (W) target to investigate hypernuclei production in small systems and at low energies. 

These reactions are very important, since they are nearby the threshold for hypermatter production. At the same time they produce hyperons with relatively low momenta with respect to the center-of-mass system. Therefore, investigating the evolution of hypermatter at low energies the study of large hypernuclei becomes possible.

In such a specific collision set up the possible reaction channels are well understood and the environment is rather clean. Generally, the reaction proceeds via the excitation of a baryonic resonance, $\pi^-+n\rightarrow \Delta ^{*,-}$ or $\pi^-+p\rightarrow \Delta ^{*,0}$ or $N^*$, with a typical mass of the Delta or $N^*$ resonance of 2 GeV. The baryon resonance (moving forward with respect to the target system) then decays after approx. 1-2 fm/c mostly into $\pi+N$ (leading to the production of protons and neutrons in the forward direction). However, the resonance may also lead to the production of $\Lambda$'s via the decay into $\Lambda+K$, and even $\Xi$ production is possible \cite{Steinheimer:2016jjk}. Further interaction of the Lambda or cascade inside the nucleus may slow down the $\Lambda/\Xi$ and allow for a binding or multi-fragmentation into a hypernucleus of varying size. The deceleration of the hyperon will of course depend on the size of the target nucleus and is more pronounced in the bigger tungsten than in the carbon target.

The pion beam measurements at HADES \cite{HADES:2017mzn} therefore open a new route to explore the properties of hypernuclei under rather well controlled and different conditions than at the LHC \cite{Donigus:2020fon}. This process may further allow to create larger hypernuclei than at the LHC (or even double strange hypernuclei) to explore the properties of strange matter which are relevant for the physics of neutron stars \cite{Ozel:2010bz,Bonanno:2011ch,Lastowiecki:2011hh,Blaschke:2015uva}.

To elucidate some of the questions above and showcase the potential of the pion beam measurements at HADES, we perform the first baseline predictions for the production of non-strange and strange clusters in $\pi^-+\mathrm{C}$ and $\pi^-+\mathrm{W}$ collisions.

% coalesence + multifragment
To this aim, we will employ a hybrid approach including the dynamical UrQMD model, the coalescence and the statistical multi-fragmentation model for the emitted coalescent clusters which can be in local equilibrium. The latter is a novel development and it avoids the energy balance limitations of the simplistic coalescence and its simple modifications, e.g. via the Wigner function approach. The coalescence model has been previously used to successfully describe the production of non-strange light nuclei, i.e. deuterons, tritons and helium and also the strange hypertriton and further strange clusters which have not yet been measured \cite{Gaebel:2020wid,Hillmann:2021zgj,Reichert:2022mek,Reichert:2023tww}. These results are compared to the results of the statistical multi-fragmentation approach which allows to produce (hyper)nuclei with large mass numbers relative to the system size \cite{Botvina:2020yfw,Buyukcizmeci:2020asf,Botvina:2022mmv,Buyukcizmeci:2023azb}.

\section{Model setup}
For this study the state-of-the-art Ultra-relativistic Quantum Molecular Dynamics model (UrQMD v3.5) \cite{Bass:1998ca,Bleicher:1999xi,Bleicher:2022kcu} is used. 
UrQMD is based on the covariant propagation of hadrons (the effective degrees of freedom) and hence provides an effective solution of the relativistic n-particle transport equation. As a QMD-type simulation, all n-particle correlations are kept track off during the system evolution. The imaginary part of the interactions is modeled via binary elastic and inelastic collisions which can excite resonances or color flux tubes which then decay or fragment into hadrons. Potential interactions, modeling the real part of the interactions, can be switched on. For the present study, however, potentials are not employed, instead a coalescence model and the Statistical Multi-fragmentation Model (SMM) \cite{Bondorf:1995ua} are used to calculate clusters. %It should be noted that the implemented potential interactions in UrQMD were shown \cite{Kireyeu:2022qmv,Coci:2023daq} to keep nucleons bound.

\subsection{Coalescence}
The production of light clusters and hypernuclei in UrQMD follows the coalescence approach, i.e. light (hyper)nuclei are formed on the kinetic freeze-out hypersurface from their constituents by coalescence. Cluster production via the coalescence mechanism has already proven to reliably describe cluster data over a broad range of collision energies and system sizes with two physics-motivated, energy independent parameters $\Delta x$ and $\Delta p$ (or relative velocity $\Delta v$). For more details about the specific implementation of coalescence at the kinetic freeze-out and for the specific parameters we refer the reader to \cite{Sombun:2018yqh,Hillmann:2021zgj} for light clusters and to \cite{Reichert:2022mek,Reichert:2023tww} for hypernuclei. 

%%%%%
\begin{figure} [t!]
    \centering
    \includegraphics[width=\columnwidth]{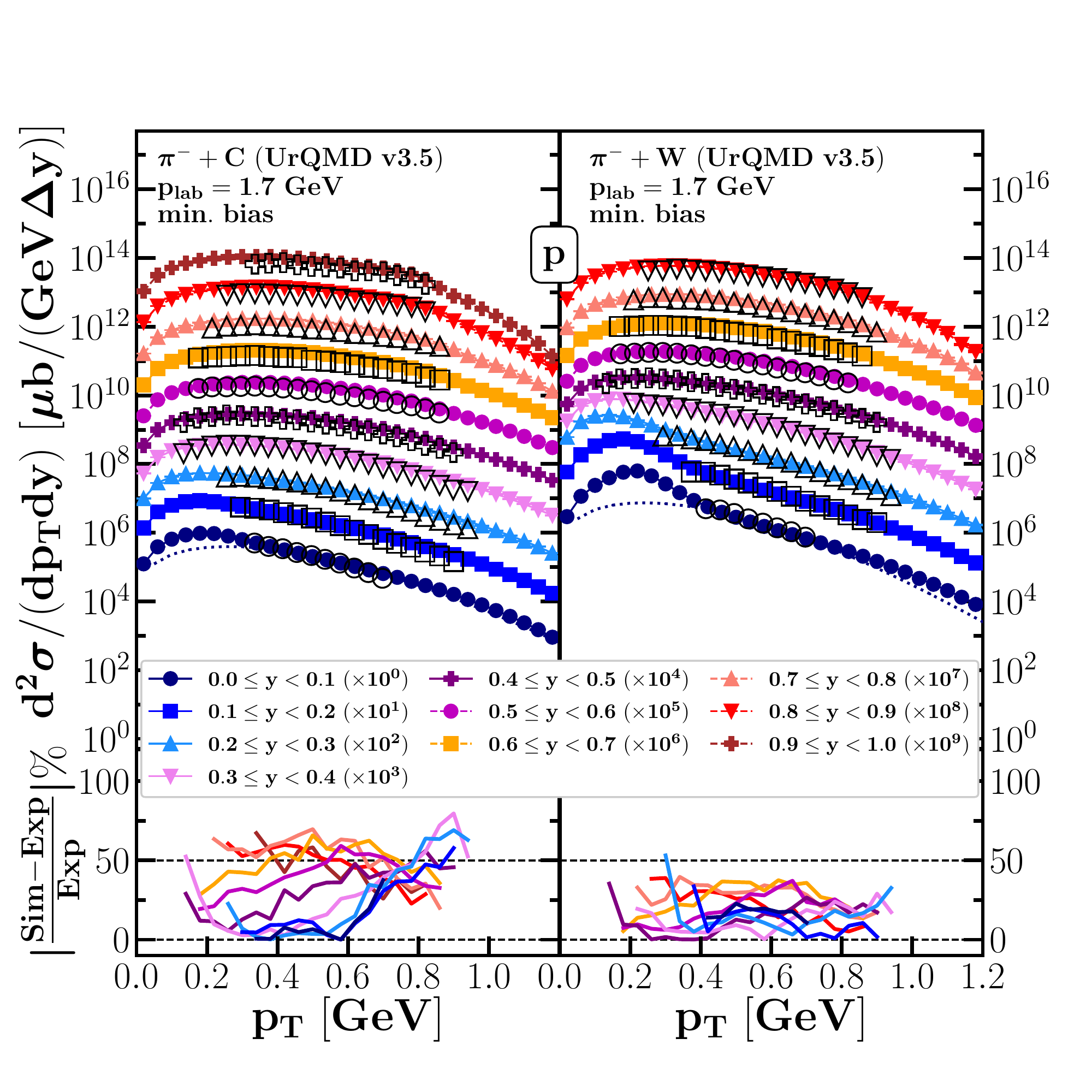}
    \caption{[Color online] Upper panel: The transverse momentum differential cross section $\mathrm{d^2}\sigma/\mathrm{d}p_\mathrm{T}\mathrm{d}y$ in $\mu$b/(GeV$\Delta y$) of free protons as a function of transverse momentum in different rapidity bins (from $0\leq y < 0.1$ to $0.9\leq y < 1.0$, the curves are successively multiplied by factors of 10 from bottom to top) for minimum bias $\pi^-+\mathrm{C}$ (left) and $\pi^-+\mathrm{W}$ (right) collisions from UrQMD (v3.5). The solid lines with symbols show the calculations and the open symbols show the recent HADES measurements \cite{HADES:2023sre}. The dotted line shows the extrapolation function used by HADES. Lower panel: The percentage error between the UrQMD simulations and the experimental data.}
    \label{fig:dsigdpt_proton}
\end{figure}

\begin{figure} [t!]
    \centering
    \includegraphics[width=\columnwidth]{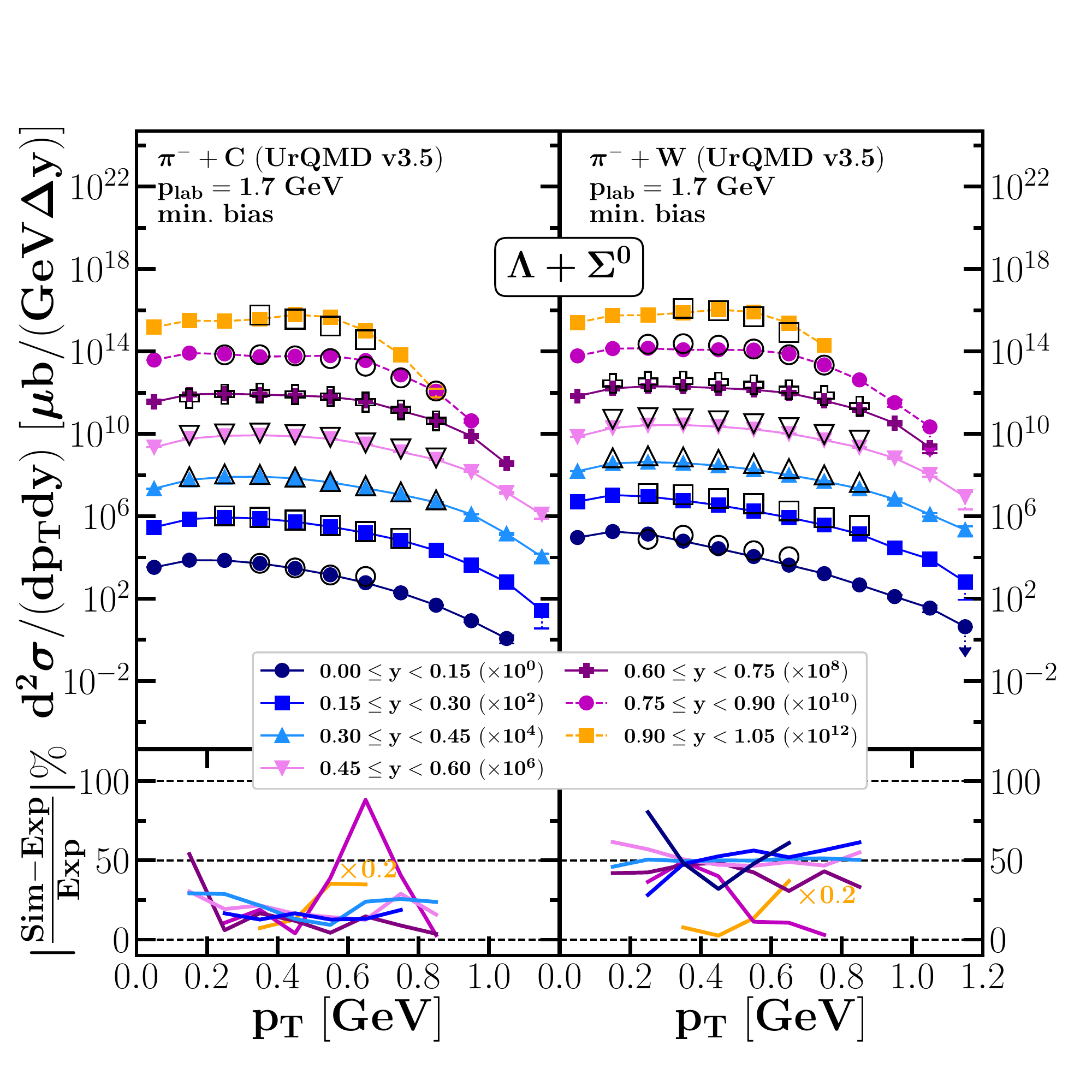}
    \caption{[Color online] Upper panel: The transverse momentum differential cross section $\mathrm{d^2}\sigma/\mathrm{d}p_\mathrm{T}\mathrm{d}y$ in $\mu$b/(GeV$\Delta y$) of $\Lambda$ as a function of transverse momentum in different rapidity bins (from $0\leq y < 0.15$ to $0.9\leq y < 1.05$, the curves are successively multiplied by factors of 100 from bottom to top) for minimum bias $\pi^-+\mathrm{C}$ (left) and $\pi^-+\mathrm{W}$ (right) collisions from UrQMD (v3.5). The solid lines with symbols show the calculations and the open symbols show the recent HADES measurements \cite{HADES:2023sre}. Lower panel: The percentage error between the UrQMD simulations and the experimental data (the last rapidity bin is scaled by a factor of $0.2$ for visibility).}
    \label{fig:dsigdpt_Lambda}
\end{figure}
%%%%

\subsection{Multi-fragmentation}
Since in $\pi+A$ reactions most of the (hyper)clusters are produced in the region of the nuclear target, we contrasted our simulations using coalescence with calculations employing the multi-fragmentation approach. Here, one assumes the formation of a larger excited nuclear system, which subsequently fragments into small parts leading to strange \cite{Botvina:2007pd} and non-strange clusters as fragmentation products. The SMM is well tested, the details of the employed approach are described in \cite{Bondorf:1995ua,Buyukcizmeci:2020asf,Botvina:2020yfw}.

\section{Results}
All results are obtained by simulating 146 million and 41 million events of minimum bias $\pi^-+\mathrm{C}$ and $\pi^-+\mathrm{W}$ collisions\footnote{
We define min. bias collisions for $\pi^-+\mathrm{C}$ with the impact parameter range $0<b<2.5$ fm and omit events without any interaction, the total cross section used for the normalization is then $\sigma^{\pi^-+\mathrm{C}}_{\rm tot}=196.35$ mb. For $\pi^-+\mathrm{W}$ we use the impact parameter range $0<b<6.5$ fm and also omit events without any interaction, the total cross section used for the normalization is then $\sigma^{\pi^-+W}_{\rm tot}=1327.32$ mb.
}
with the UrQMD model in version 3.5. All calculations are done in the target rest frame (laboratory frame in the HADES experiment). The rapidity variable $y$ thus refers to the target rest frame, i.e. $y_\mathrm{target}=0$. We note that in order to be compared to the experimental data, the free $\Lambda$'s always include the $\Sigma^0$'s as the latter ones decay into the Lambda and can not be distinguished in the experiment from initial $\Lambda$'s. However, the coalescence and multi-fragmentation routines only consider initial $\Lambda$'s as the decay of the $\Sigma^0$ happens far away due to its comparatively long life time. Further $\Sigma^0$'s do not form hypernuclei (or only barely) and do not contribute to the production of hypertritons. In the present simulation protons do not include feed down from the $\Lambda$ decay. The units used for $\mathrm{d^2}\sigma/\mathrm{d}p_\mathrm{T}\mathrm{d}y$ are $\mu$b/(GeV$\Delta y$) and for $\mathrm{d}\sigma/\mathrm{d}y$ it is $\mu$b/$\Delta y$ which are chosen in line with the presented HADES measurements \cite{HADES:2023sre}. The coalescence results originating from UrQMD are evaluated at the respective kinetic freeze-out of the particles, while the SMM employs input from the UrQMD at a fixed time which is set to 20 fm/c in the present analysis. Furthermore all particles (including those which are identified as spectators by UrQMD) are given to the statistical multi-fragmentation approach. The coalescence parameter for the primary hot cluster recognition in the SMM model has been set to $v_c=0.22$ as this value has been shown to provide good results \cite{Botvina:2014lga,Buyukcizmeci:2020asf,Botvina:2020yfw}.

\subsection{Transverse momentum spectra of protons and Lambda hyperons}
\label{subsec:pTproton}
We start the investigation by inspecting the double differential transverse momentum spectra of the protons and the $\Lambda$'s. 

Fig. \ref{fig:dsigdpt_proton} shows the transverse momentum differential cross section $\mathrm{d^2}\sigma/\mathrm{d}p_\mathrm{T}\mathrm{d}y$ in $\mu$b/(GeV$\Delta y$) of protons as a function of transverse momentum in different rapidity bins (from $0\leq y < 0.1$ to $0.9\leq y < 1.0$, the curves are successively multiplied by factors of 10 from bottom to top) for minimum bias $\pi^-+\mathrm{C}$ (left panel) and $\pi^-+\mathrm{W}$ (right panel) collisions from UrQMD (v3.5). The solid lines with symbols show the calculations. The open symbols show the recent HADES measurements \cite{HADES:2023sre} and the dotted line shows the exponential curve\footnote{Exponential function parameterized by the yield integral and slope parameter $\sim  C(y)p_\mathrm{T} \sqrt{p_\mathrm{T}^2+m_{0}^2} \exp{\left[-\sfrac{\sqrt{p_\mathrm{T}^2+m_{0}^2}}{T(y)}\right]}$ in line with the HADES analysis \cite{HADES:2023sre}.} fitted to the simulation data in the same $p_\mathrm{T}$ acceptance as in the HADES setup. We chose to show the exponential fit of the transverse momentum spectra only in the rapidity window $0\leq y <  0.1$ for brevity in the plot. The percentage error between the UrQMD simulations and the experimental data is shown in the lower panel. 

One observes that the UrQMD model calculations of the differential cross section for protons agree qualitatively well with the experimental data \cite{HADES:2023sre} in the acceptance region in both collision systems and over all investigated rapidity bins. The slope parameters in the model and the data agree well. A very prominent effect in the proton distributions is noticeable at low rapidity: The UrQMD simulation shows more protons with relatively low transverse momenta and at low rapidity than from the exponential fit to the data, as expected. These are mostly spectator-like nucleons which will eventually form large excited residues decaying into fragments on the later reaction stages. Unfortunately, due to the acceptance of the HADES detector, the phase space region in which the residual nucleus is located (i.e. around zero rapidity and small transverse momenta) is not fully covered by the experiment. Thus, the experiment cannot directly observe the residue free protons at transverse momenta $p_\mathrm{T} \leq 0.4$~GeV around zero rapidity experimentally. The effect is apparent for both systems and becomes stronger as the system size increases. We will see in the next section that the use of experimental extrapolation results in slightly different rapidity densities near $y \approx 0$.

% Figure \ref{fig:dsigdpt_proton} (lower panels) presents the percentage differences in $p_\mathrm{T}$ spectra between the UrQMD (v3.5) model and the HADES experimental data \cite{HADES:2023sre}  for protons. These differences remain below $50\%$ across the $p_\mathrm{T}$ range and all rapidity bins. In the carbon system, the deviations becomes stronger with increasing momentum and further window. Conversely, in the tungsten system, the deviations indicate less dependence on both $p_\mathrm{T}$ and rapidity.

Fig. \ref{fig:dsigdpt_Lambda} shows the transverse momentum differential cross section $\mathrm{d^2}\sigma/\mathrm{d}p_\mathrm{T}\mathrm{d}y$ in $\mu$b/(GeV$\Delta y$) of $\Lambda$'s as a function of transverse momentum in different rapidity bins (from $0\leq y < 0.15$ to $0.9\leq y < 1.05$, the curves are successively multiplied by factors of 100 from bottom to top) for minimum bias $\pi^-+\mathrm{C}$ (left panel) and $\pi^-+\mathrm{W}$ (right panel) collisions from UrQMD (v3.5). The solid lines with symbols show the calculations and the open symbols show the recent HADES measurements \cite{HADES:2023sre}. The deviation between the UrQMD simulations and experimental data is shown in the lower panel.

The UrQMD model calculations agree well with the data across the transverse momentum range in all rapidity bins with the sharp drop-off behaviors at high 
$p_\mathrm{T}$ values. This drop-off in the transverse momentum spectrum towards higher rapidities arises from the limited available energy in the collision.

\subsection{Rapidity distributions of protons and Lambda hyperons}
\begin{figure} [t]
    \centering
    \includegraphics[width=\columnwidth]{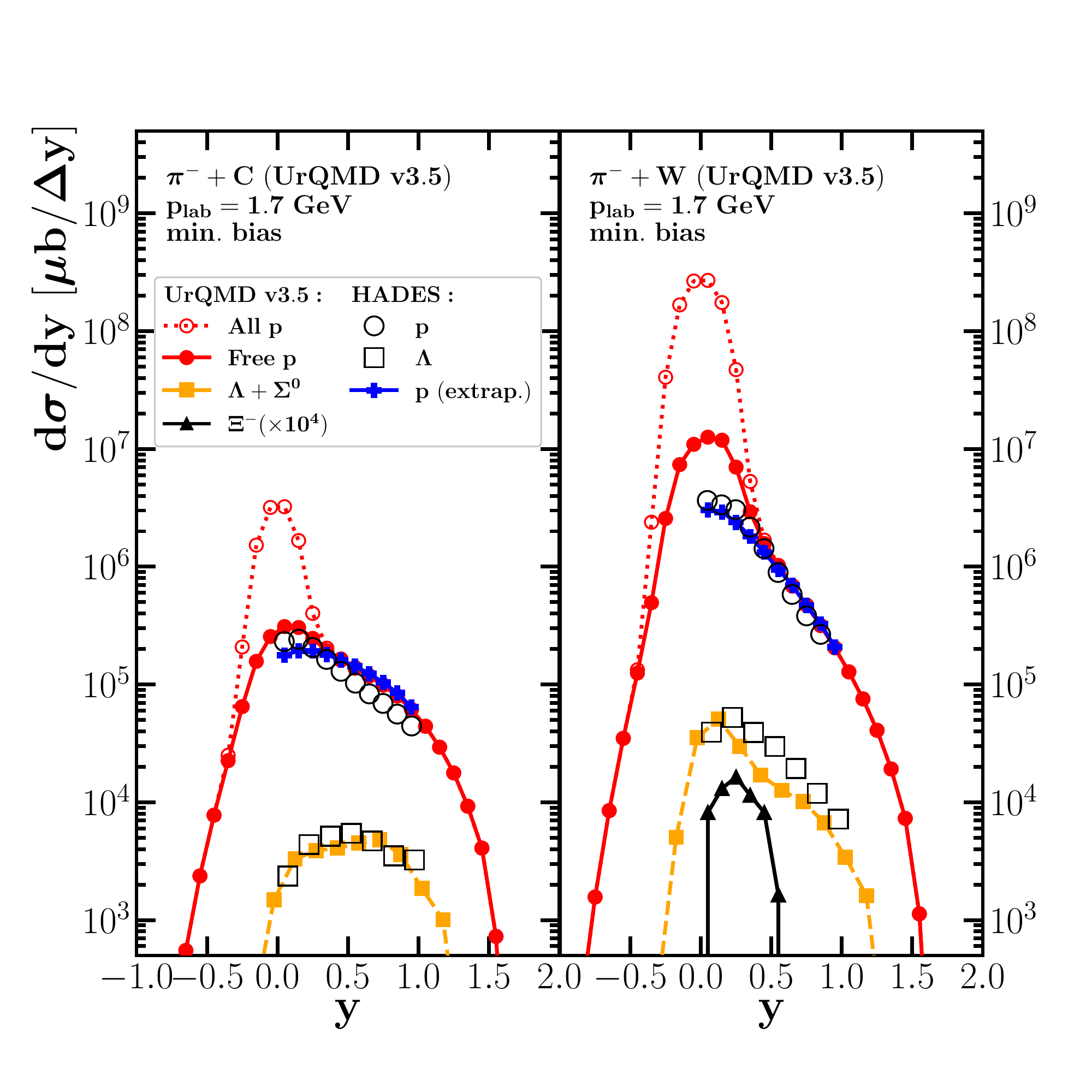}
    \caption{[Color online] The rapidity differential cross section $\mathrm{d}\sigma/\mathrm{d}y$ in $\mu$b/$\Delta y$ of protons (red), $\Lambda$'s (orange), and $\Xi$'s (black) as a function of the rapidity for minimum bias $\pi^-+\mathrm{C}$ (left panel) and $\pi^-+\mathrm{W}$ (right panel) collisions. The results from UrQMD are shown as colored lines with symbols, while the open black symbols without lines depict the recent HADES measurements \cite{HADES:2023sre}.The blue crosses show the result using the experimental fit function for the $p_\mathrm{T}$ extrapolation.}
    \label{fig:dsigdy}
\end{figure}

Fig. \ref{fig:dsigdy} shows the rapidity differential cross section $\mathrm{d}\sigma/\mathrm{d}y$ in $\mu$b/$\Delta y$ of protons (red) and $\Lambda$'s (orange) as a function of the rapidity for minimum bias $\pi^-+\mathrm{C}$ (left panel) and $\pi^-+\mathrm{W}$ (right panel) collisions. The results from UrQMD are shown as colored lines with symbols and open black symbols without lines depict the recent HADES measurements \cite{HADES:2023sre}.

The discussion begins with the rapidity spectra of protons. Firstly the free protons obtained through UrQMD (solid red line with full circles) should be compared to all protons (including those bound in clusters via coalescence, shown as open red circles with dotted lines). One clearly observes that at forward and backward rapidities the distributions match each other and describe the HADES data (shown as black open symbols) very well. Only in the target rapidity region the calculated proton spectra differ. The difference is obviously accounted to cluster formation as well as to spectator identification. However, there is a remaining difference between the free protons from the simulation and the data. This difference can be traced back to the extrapolation used in the experiment towards low $p_\mathrm{T}$ (Section \ref{subsec:pTproton} and Fig. \ref{fig:dsigdpt_proton}). Using the same extrapolation as in the experimental data leads to the blue line with the crosses, which matches the experimental data very well and fully resolves the tension between the calculation and the data in the target region. %Moving on to the free protons obtained via the SMM (shown by purple lines) one observes that the fragmentation calculation catches up with the transport calculation at forward and backward rapidities and also described the empirical data well. In the target rapidity region the SMM model is in between the UrQMD curves employing all protons and free protons. This comes from the fact that the spectator\footnote{Spectator in this collision setup doesn't refer to nucleons bypassing the collision zone, but rather to nucleons not participating in the imaginary part of the interactions.} identification is ill-defined. Both models, however, show a contrast to the free protons (shown as a blue line with symbols) obtained by following the experimental extraction technique, i.e. fitting the transverse momentum spectra with a Boltzmann distribution and extracting the $\mathrm{d}\sigma/\mathrm{d}y$, particularly near the target region. The explanation behind this is, as previously noted, the integral of the transverse momentum spectrum cannot describe the residue free protons around zero rapidity in both system. 
We therefore suggest that future investigations adjust their fit function to account for these residue free protons. %Furthermore, we also display the spectra of all protons (dashed red lines with open symbols). These all protons data can be obtained by inclusion of bounded protons in the cluster formation, which can be directly detected by the experiment, to the free protons. 

The measured $\Lambda$ distribution in the  $\pi^-+\mathrm{C}$ collision system is well described by the UrQMD model calculations, while in the larger $\pi^-+\mathrm{W}$ system, the Lambda production in forward direction is slightly underestimated as compared to the experimental data. However, the general trend and multiplicity is captured well providing a reasonable description of the proton and Lambda distributions in longitudinal and transverse direction for both collision systems which allows to investigate light and hypercluster production in further detail.

Let us finally note that even in at such low energy also the production of $\Xi$ baryons is possible (black triangles) in the large tungsten system with a cross section on the order of a few $\mu b$. This may open the exciting possibility to explore multistrange hyperclusters similar to the approach suggested in \cite{Pochodzalla:2011rz} although without the need for an anti-proton beam.

\begin{figure} [t]
    \centering
    \includegraphics[width=\columnwidth]{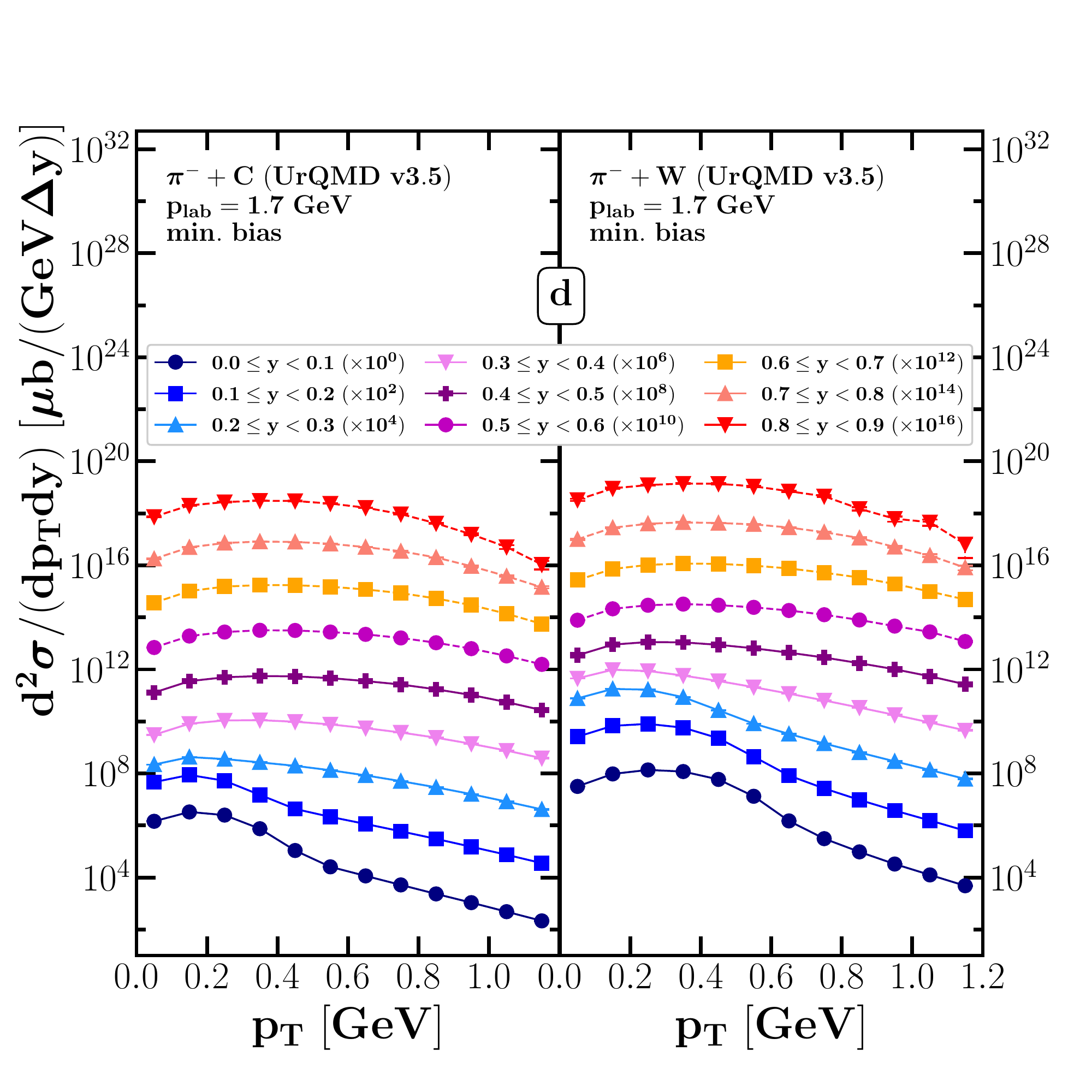}
    \caption{[Color online] The transverse momentum differential cross section $\mathrm{d^2}\sigma/\mathrm{d}p_\mathrm{T}\mathrm{d}y$ in $\mu$b/(GeV$\Delta y$) of deuterons as a function of transverse momentum in different rapidity bins (from $0\leq y < 0.1$ to $0.8\leq y < 0.9$, the curves are successively multiplied by factors of 100 from bottom to top) for minimum bias $\pi^-+\mathrm{C}$ (left panel) and $\pi^-+\mathrm{W}$ (right panel) collisions from UrQMD.}
    \label{fig:dsigdpt_d}
\end{figure}
\begin{figure} [t]
    \centering
    \includegraphics[width=\columnwidth]{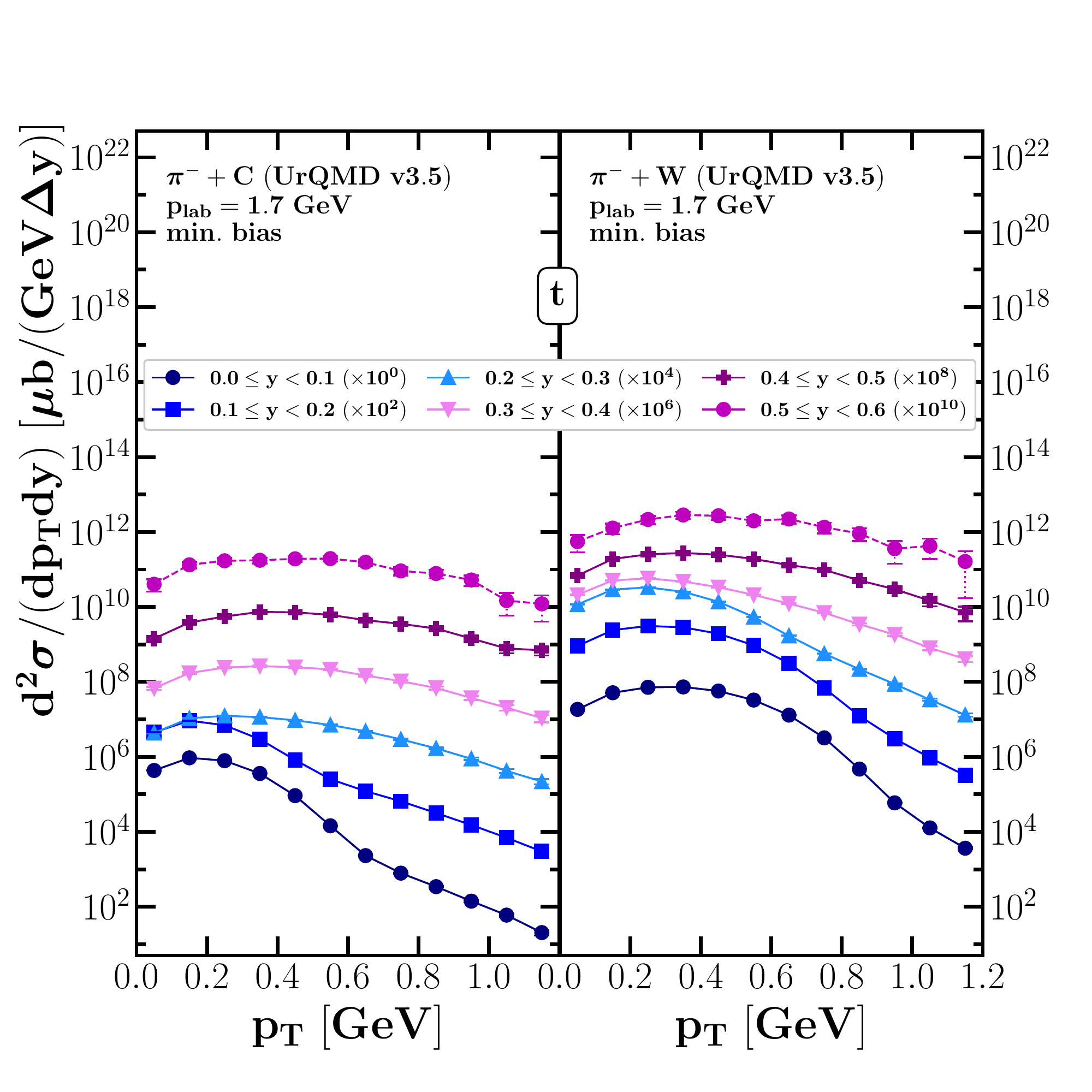}
    \caption{[Color online] The transverse momentum differential cross section $\mathrm{d^2}\sigma/\mathrm{d}p_\mathrm{T}\mathrm{d}y$ in $\mu$b/(GeV$\Delta y$) of tritons as a function of transverse momentum in different rapidity bins (from $0\leq y < 0.1$ to $0.5\leq y < 0.6$, the curves are successively multiplied by factors of 100 from bottom to top) for minimum bias $\pi^-+\mathrm{C}$ (left panel) and $\pi^-+\mathrm{W}$ (right panel) collisions from UrQMD.}
    \label{fig:dsigdpt_t}
\end{figure}
\begin{figure} [t]
    \centering
    \includegraphics[width=\columnwidth]{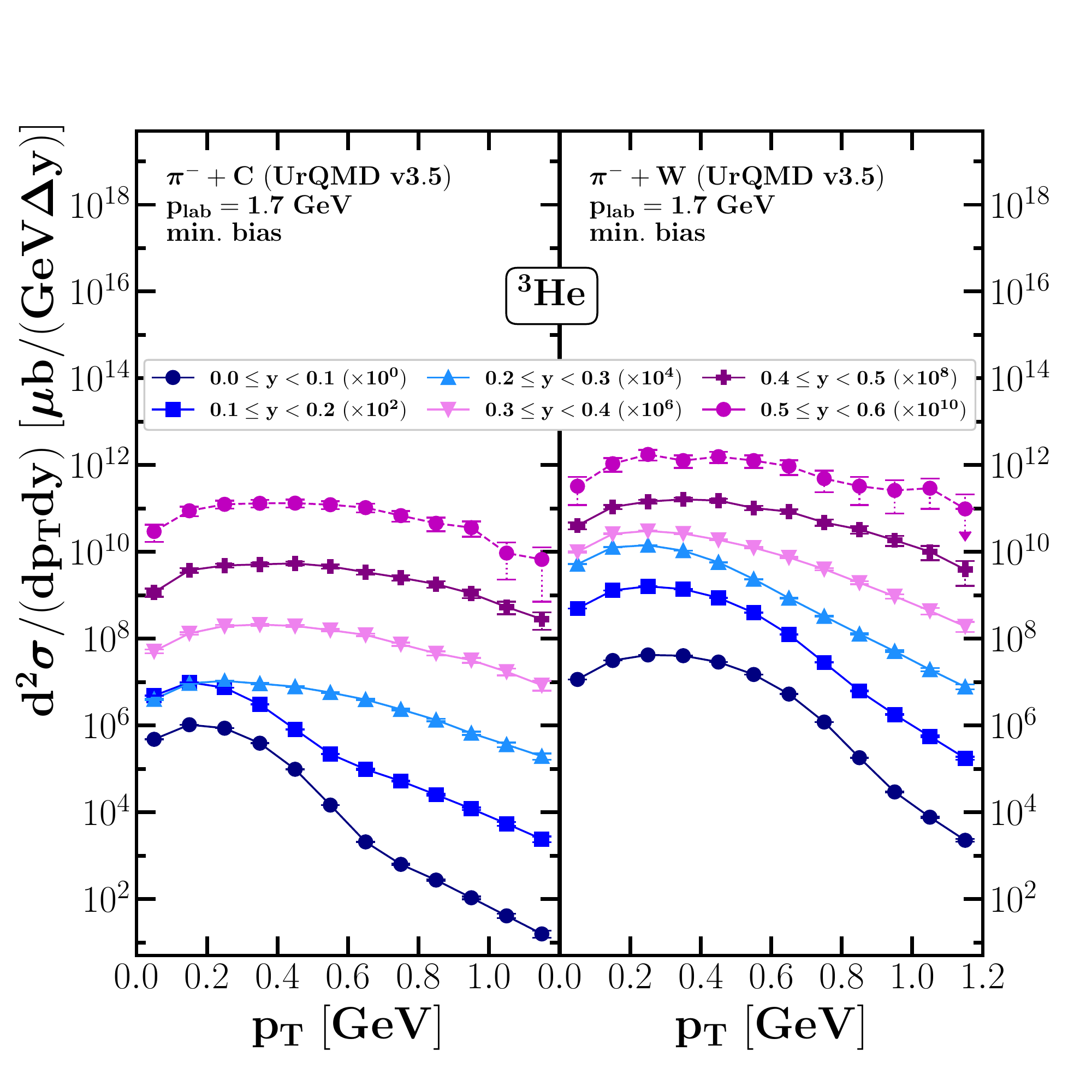}
    \caption{[Color online] The transverse momentum differential cross section $\mathrm{d^2}\sigma/\mathrm{d}p_\mathrm{T}\mathrm{d}y$ in $\mu$b/(GeV$\Delta y$) of $^3$He as a function of transverse momentum in different rapidity bins (from $0\leq y < 0.1$ to $0.5\leq y < 0.6$, the curves are successively multiplied by factors of 100 from bottom to top) for minimum bias $\pi^-+\mathrm{C}$ (left panel) and $\pi^-+\mathrm{W}$ (right panel) collisions from UrQMD.}
    \label{fig:dsigdpt_3he}
\end{figure}
\begin{figure} [t]
    \centering
    \includegraphics[width=\columnwidth]{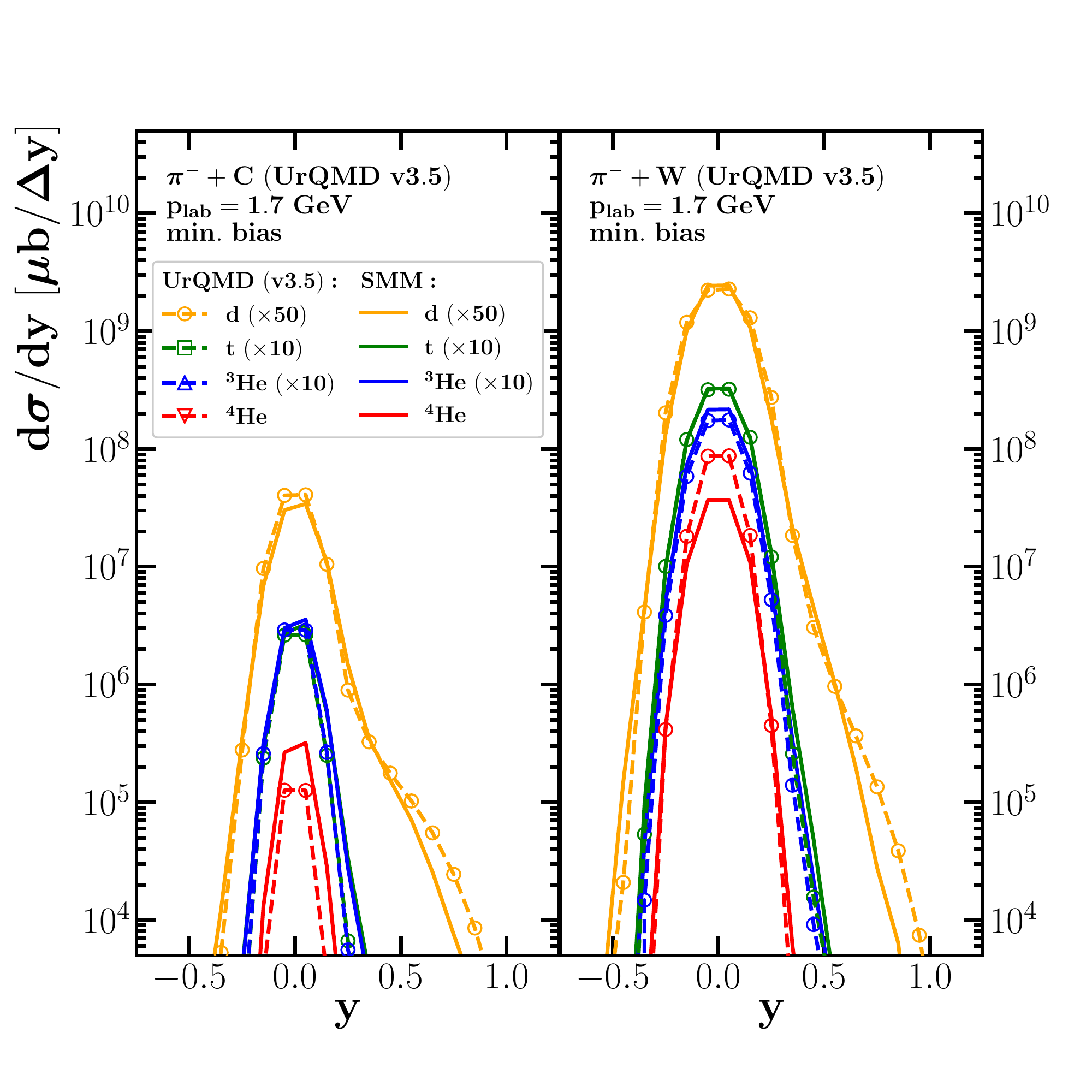}
    \caption{[Color online] The rapidity differential cross section $\mathrm{d}\sigma/\mathrm{d}y$ in $\mu$b/$\Delta y$ of deuterons (orange), tritons (green), $^3$He (blue) and $^4$He (red) as a function of the rapidity for minimum bias $\pi^-+\mathrm{C}$ (left panel) and $\pi^-+\mathrm{W}$ (right panel) collisions from UrQMD (v3.5) as denoted by dashed lines with symbols and from statistical multi-fragmentation model (SMM) as denoted by solid lines without symbols.}
    \label{fig:dsigdy_lightcluster}
\end{figure}

\subsection{Transverse momentum spectra and rapidity distribution of light clusters}
After having benchmarked the proton and $\Lambda$ spectra, we are now ready to address cluster production.

In Fig. \ref{fig:dsigdpt_d} the transverse momentum differential cross section $\mathrm{d^2}\sigma/\mathrm{d}p_\mathrm{T}\mathrm{d}y$ in $\mu$b/(GeV$\Delta y$) of deuterons as a function of transverse momentum in different rapidity bins (from $0\leq y < 0.1$ to $0.8\leq y < 0.9$) are shown for minimum bias $\pi^-+\mathrm{C}$ (left panel) and $\pi^-+\mathrm{W}$ (right panel) collisions from UrQMD. The curves are successively multiplied by factors of 100 from bottom to top for better visibility.

Next, we present in Fig. \ref{fig:dsigdpt_t}  the transverse momentum differential cross section $\mathrm{d^2}\sigma/\mathrm{d}p_\mathrm{T}\mathrm{d}y$ in $\mu$b/(GeV$\Delta y$) of tritons as a function of transverse momentum in different rapidity bins (from $0\leq y < 0.1$ to $0.5\leq y < 0.6$, the curves are successively multiplied by factors of 100 from bottom to top) for minimum bias $\pi^-+\mathrm{C}$ (left panel) and $\pi^-+\mathrm{W}$ (right panel) collisions from UrQMD.

Finally, Fig. \ref{fig:dsigdpt_3he} shows the transverse momentum differential cross section $\mathrm{d^2}\sigma/\mathrm{d}p_\mathrm{T}\mathrm{d}y$ in $\mu$b/(GeV$\Delta y$) of $^3$He as a function of transverse momentum in different rapidity bins (from $0\leq y < 0.1$ to $0.5\leq y < 0.6$, the curves are successively multiplied by factors of 100 from bottom to top) for minimum bias $\pi^-+\mathrm{C}$ (left panel) and $\pi^-+\mathrm{W}$ (right panel) collisions from UrQMD.

Generally, we observe a substantial amount of cluster production, especially in the target rapidity region. 

We summarize our results in Fig. \ref{fig:dsigdy_lightcluster} showing a comparison for the rapidity differential cross section $\mathrm{d}\sigma/\mathrm{d}y$ in $\mu$b/$\Delta y$ of deuterons (orange), tritons (green), $^3$He (blue), and $^4$He (red) as a function of the rapidity for minimum bias $\pi^-+\mathrm{C}$ (left panel) and $\pi^-+\mathrm{W}$ (right panel) collisions from UrQMD denoted as dashed lines with open symbols. We further compare yields from the UrQMD coalescence approach to the SMM yields denoted by solid lines without symbols.

The peaks of the distributions are centered around target rapidity (i.e. where the residual nucleus is located and fragments). Towards forward rapidity the deuterons show a moderate decrease while the tritons, Helium-3, and Helium-4 show a rapid decline. This is due to the fact that it is much less likely for the impinging $\pi^-$ to detach 3 or more nucleons from the target than 2 nucleons. We further observe I) a broadening of the distributions around the peak and II) a protrusion of the distribution of deuterons and the other investigated nuclei from the peak around the target rapidity in both systems. The physical explanation of I) is that the broadening is a consequence of the fact that both the coalescence and the multi-fragmentation processes contribute essentially to the fragment formation around the target region. II) The protruding part of the deuterons in the carbon system is more discernible than in the tungsten system, indicating a lower stopping power in the smaller carbon system as compared to the tungsten system.

Also, the cluster yields from coalescence (UrQMD) and multi-fragmentation (SMM) agree very well with each other, which will allow us to use the SMM to extrapolate to high mass hyperclusters.

\subsection{Transverse momentum spectra and rapidity distribution of hypernuclei}
Finally, we turn towards the most exciting physics possibilities of the pion beam experiment, namely the production of hypernuclei, and especially the hypertriton. 

\begin{figure} [t]
    \centering
    \includegraphics[width=\columnwidth]{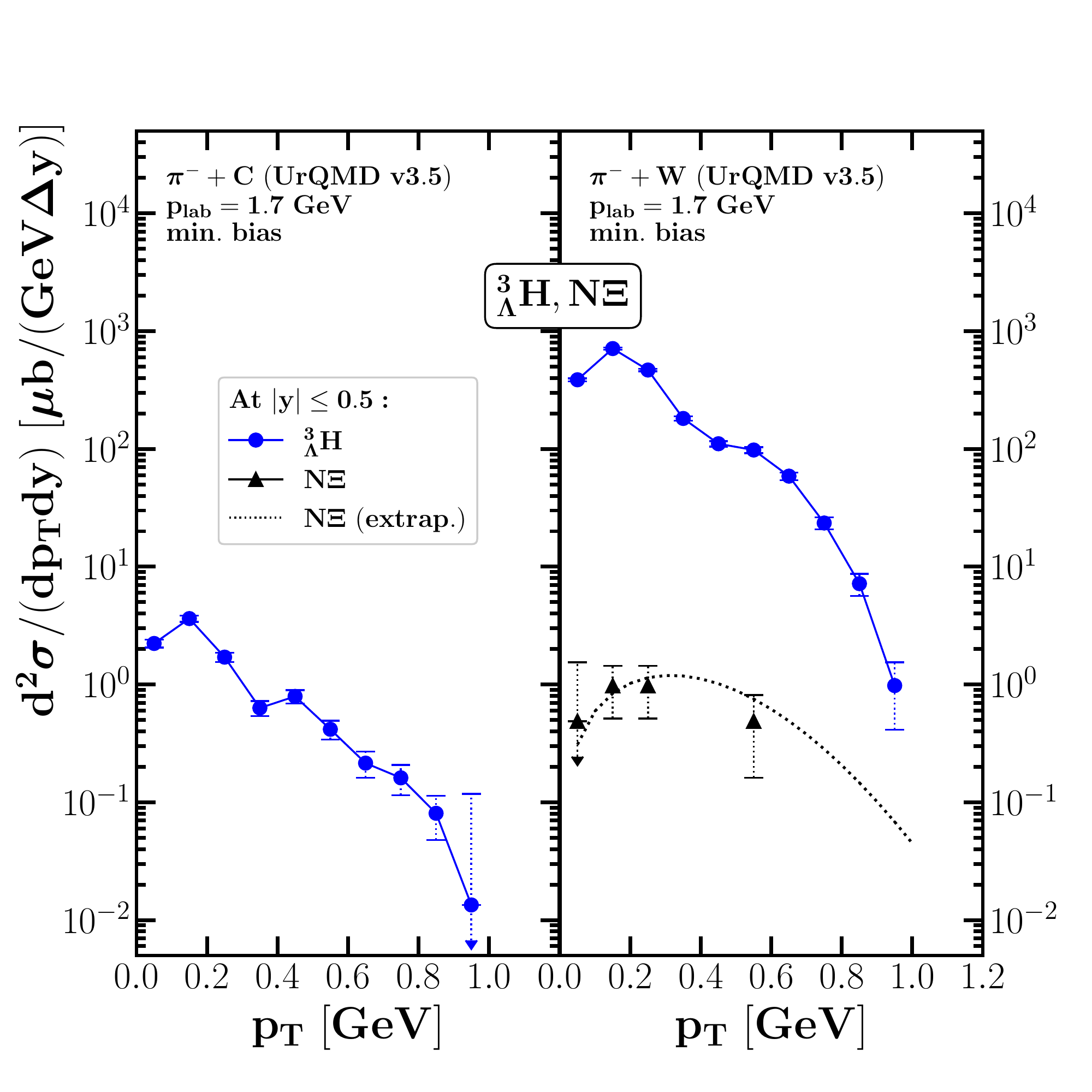}
    \caption{[Color online] The transverse momentum differential cross section $\mathrm{d^2}\sigma/\mathrm{d}p_\mathrm{T}\mathrm{d}y$ in $\mu$b/(GeV$\Delta y$) of $^3_\Lambda$H (blue) and $N\Xi$ (black) as a function of transverse momentum at $|y| \leq 0.5$, for minimum bias $\pi^-+\mathrm{C}$ (left panel) and $\pi^-+\mathrm{W}$ (right panel) collisions from UrQMD. The extrapolated $N\Xi$ is shown as a dotted line.}
    \label{fig:dsigdpt_3LH}
\end{figure}

Fig. \ref{fig:dsigdpt_3LH} shows the transverse momentum differential cross section $\mathrm{d^2}\sigma/\mathrm{d}p_\mathrm{T}\mathrm{d}y$ in $\mu$b/(GeV$\Delta y$) of the hypertriton $^3_\Lambda$H and the speculated $N\Xi$ clusters as a function of transverse momentum at midrapidity $|y|\leq 0.5$ for minimum bias $\pi^-+\mathrm{C}$ (left panel) and $\pi^-+\mathrm{W}$ (right panel) collisions from UrQMD.

The main observation from the predicted transverse momentum spectra is that the $^3_\Lambda$H is copiously produced with transverse momenta accessible within the HADES acceptance, while $N\Xi$ clusters maybe in reach.

\begin{figure} [t!]
    \centering
    \includegraphics[width=\columnwidth]{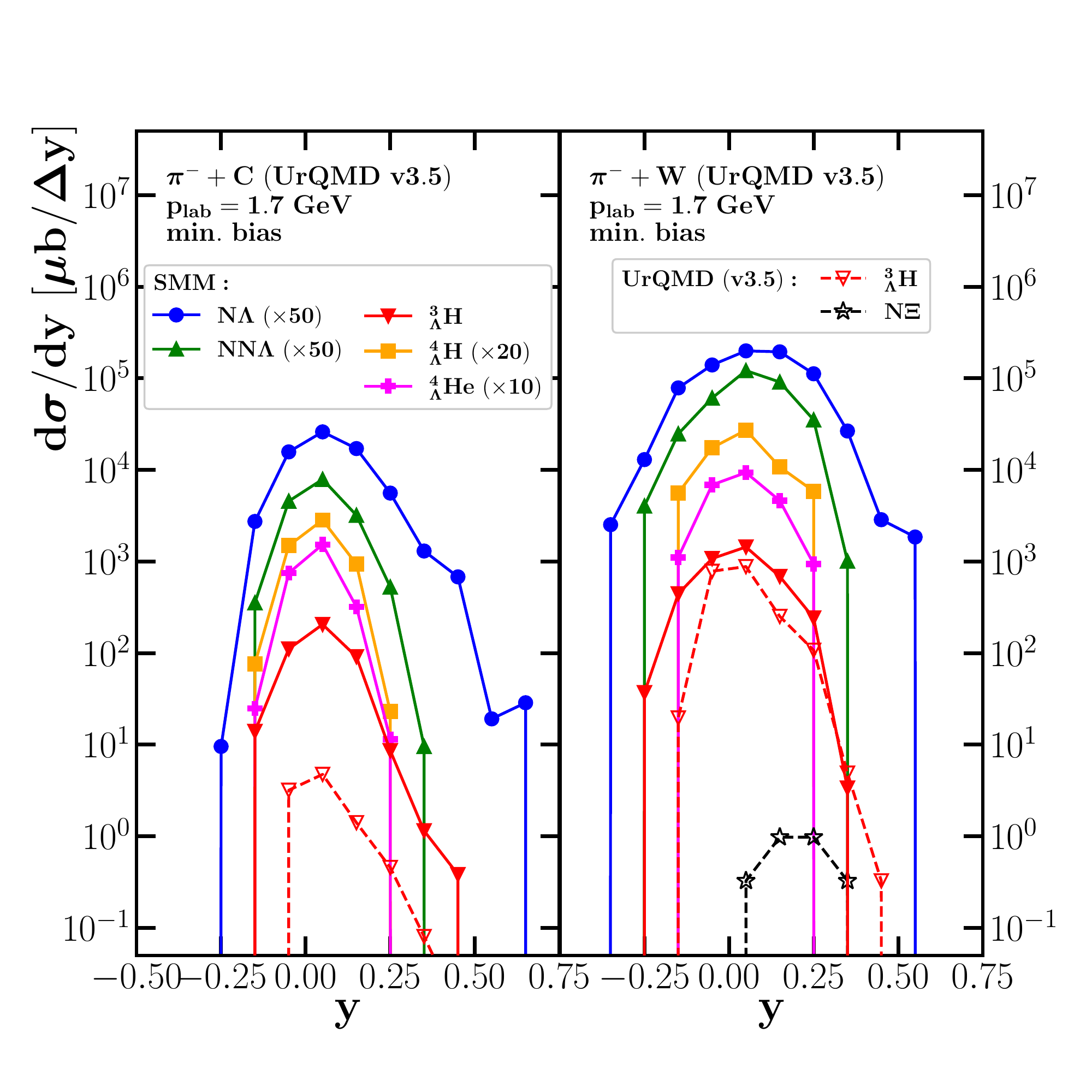}
    \caption{[Color online] The rapidity differential cross section $\mathrm{d}\sigma/\mathrm{d}y$ in $\mu$b/$\Delta y$ of the $N\Lambda$ (blue), $NN\Lambda$ (green), $^3_\Lambda$H (red), $^4_\Lambda$H (yellow), $^4_\Lambda$He (pink), and $N\Xi$ (black) as a function of the rapidity for minimum bias $\pi^-+\mathrm{C}$ (left panel) and $\pi^-+\mathrm{W}$ (right panel) collisions from UrQMD with coalescence (v3.5) denoted as dashed lines with open symbols and statistical multi-fragmentation model (SMM) denoted as solid lines with full symbols.}
    \label{fig:dsigdy_hypernuclei}
\end{figure}

Finally, we combine the different transverse momentum distributions and obtain in Fig. \ref{fig:dsigdy_hypernuclei} the rapidity differential cross section $\mathrm{d}\sigma/\mathrm{d}y$ in $\mu$b/$\Delta y$ of the $^3_\Lambda$H (dashed red lines with open symbols), $N\Xi$ (black stars) as well as other clusters as a function of the rapidity for minimum bias $\pi^-+\mathrm{C}$ (left panel) and $\pi^-+\mathrm{W}$ (right panel) collisions from UrQMD with coalescence denoted by dashed lines and from SMM denoted by solid lines with full symbols.

As expected, the rapidity density of the cross section of $^3_\Lambda$H production has a peak value in the target rapidity region. The yield in the $\pi^-+\mathrm{W}$ system is substantially larger than in the $\pi^-+\mathrm{C}$ system, due to a higher hyperon multiplicity and due to the larger stopping power of the bigger nucleus, which increases the formation probability. Turning cross section into numbers, this suggests that $\mathcal{O}(10^{-3})$ hypertritons per event can be expected in these collisions. This means that measured statistics reported by the HADES collaboration (approx. $10^8$ recorded events for each system) in Ref. \cite{HADES:2023sre} is sufficient to extract up to $\mathcal{O}(10^5)$ $^3_\Lambda$H from their full data set allowing for a detailed and highly differential measurement of the hypertriton.

The light cluster and hypertriton formation from UrQMD and SMM are agreement for the larger system, however for the smaller system they differ by a factor of 10, which may indicate the systematic error for the hypertriton production. Fig. \ref{fig:dsigdy_hypernuclei} therefore additionally shows the rapidity densities of the $N\Lambda$ (blue line with circles), the $NN\Lambda$ (green line with triangles-up), the ${}^4_\Lambda$H (yellow line with squares) and the ${}^4_\Lambda$He (pink line with crosses) as calculated with the SMM model. One observes that these hypernuclei are also produced copiously and confirms that the stopping power from the target nucleus allows only a few nucleons to fly in the forward rapidity direction. As a result, the $N\Lambda$ (blue) formation with $A=2$ extends more toward forward rapidity compare to other hypernuclei with $A\geq 3$.


\subsection{Light and hyperfragments of larger mass numbers}
Within the statistical multi-fragmentation model also larger light nuclei and hypernuclei are formed due to the statistical decay of the residues and excited coalescent clusters. This allows for the first time to estimate the total abundance of $\Lambda$-hypernuclei in the analyzed $\pi^-+\mathrm{C/W}$ systems at 1.7 GeV incident momentum.  

\begin{figure} [t!]
    \centering
    \includegraphics[width=\columnwidth]{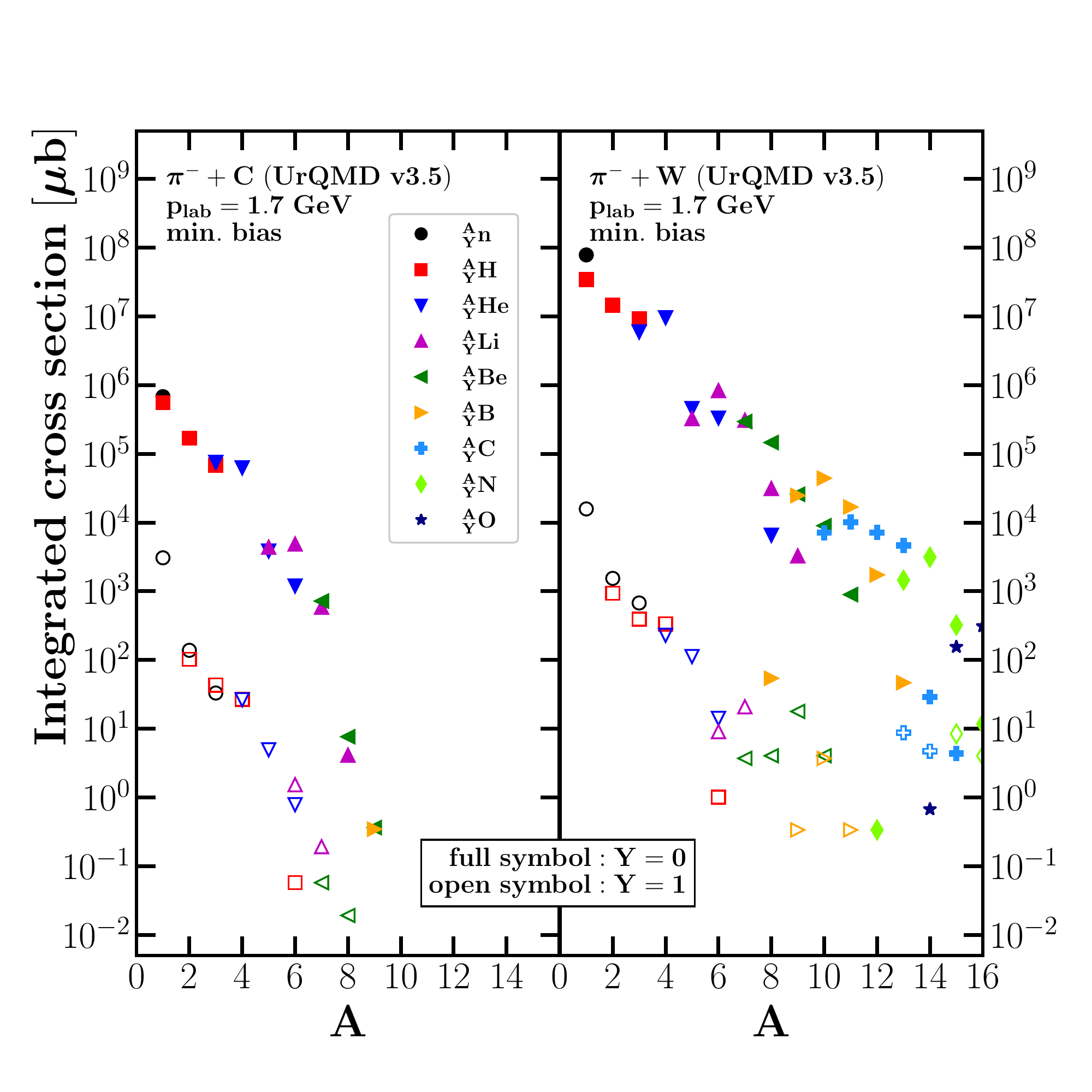}
    \caption{[Color online] The integrated cross section of light nuclei (full symbols) and hypernuclei (single-strange as open symbols) production with different charges $Z$ (denoted by the color) as a function of their mass number $A$ for minimum bias $\pi^-+\mathrm{C}$ and $\pi^-+\mathrm{W}$ collisions from a statistical multi-fragmentation (SMM) analysis of the UrQMD (v3.5) data.}
    \label{fig:multiplicty_allnuclei}
\end{figure}

Fig. \ref{fig:multiplicty_allnuclei} shows the integrated cross section of light nuclei (full symbols) and hypernuclei (single-strange as open symbols) production with different charges $Z$ (denoted by the color) as a function of their mass number $A$ for minimum bias $\pi^-+\mathrm{C}$ and $\pi^-+\mathrm{W}$ collisions from a statistical multi-fragmentation (SMM) analysis of the UrQMD (v3.5) data.
Although the light cluster and hypernuclei abundances fall off approximately exponentially with increasing mass number, a phenomenon also observed in the
production of light nuclei at RHIC \cite{STAR:2011eej}, their integrated production cross section might still allow for a signal extraction by the HADES collaboration or in an upcoming $\pi+A$ experiment with higher beam luminosity. As light nuclear fragments with a mass numbers up to $A=16$ are still produced with a yield of $10^{-6}$ per event and hypernuclei with mass number $A=10$ with a yield of $10^{-7}$ per event, this allows for the first time to study light clusters and hypernuclei with large $A$ in $\pi A$ collisions. As reported by the HADES collaboration in \cite{Ardid:1999mqs} their pion beam can reach beam momenta of up to $p_\mathrm{beam}=2.5$ GeV lifting the abundances drastically and allowing for significant $\Xi$-hypernuclei production as well. This may offer the unique opportunity to implement parts of the hypermatter physics program proposed for PANDA in the HADES $\pi+A$ program.

\section{Conclusion}
In this paper we have employed the state-of-the-art Ultra-relativistic Quantum Molecular Dynamics (UrQMD) microscopic transport model to calculate $\pi^-+\mathrm{C}$ and $\pi^-+\mathrm{W}$ collisions at p$_\mathrm{lab}=1.7$ GeV. We found good agreement between the calculated and measured transverse momentum and rapidity distributions of protons and $\Lambda$'s. An important aspect for the description of the proton spectra in the target region was the inclusion of cluster production to obtain the free proton spectra using the coalescence approach. We have further contrasted our results on light nuclei and hypernuclei production from a coalescence approach with a Statistical Multi-fragmentation Model (SMM). We have predicted the light cluster (deuteron, triton and Helium-3) double differential cross sections and rapidity densities using the coalescence mechanism and SMM model. Finally, we applied the coalescence mechanism and the SMM to predict hypernuclei double differential cross sections. Both non-strange and strange cluster yields are sufficiently high to be accessible by the HADES collaboration's statistics. 

As a final remark, the potential of double strange hypernuclei production is discussed. It was suggested that $\Xi$ production proceeds also via the formation of a resonance that decays via $N^*\rightarrow \Xi+K+K$ \cite{Steinheimer:2015sha}. If such a resonance exists, then the HADES pion beam set-up with a selection of pions with higher momenta (reaching at least a center-of-mass energy of $\sqrt s = m_\Xi + 2m_K$) would allow to perform similar studies as with the PANDA detector for multi-strange hyperfragments. In the same manner, the double-$\Lambda$ hypernuclei can be produced via $\Xi+N+N \leftrightarrow \Lambda+\Lambda+N$.

\section*{Acknowledgments}
The authors thank Manuel Lorenz, Christoph Blume and Benjamin D\"onigus for fruitful discussion about light and hypernuclei. 
This article is part of a project that has received funding from the European Union’s Horizon 2020 research and innovation programme under grant agreement STRONG – 2020 - No 824093.
This work was supported by DAAD (PPP Thailand) and (PPP Turkey).
The computational resources for this project were provided by the Center for Scientific Computing of the GU Frankfurt and the Goethe--HLR.
N. B. acknowledges the Scientific and Technological Research Council of Turkey
(TUBITAK) support under Project No. 121N420.
Also this research has received funding support from the NSRF via the Program Management Unit for Human Resources \& Institutional Development, Research and Innovation [grant number B16F640076].

%\bibliography{refs}% Produces the bibliography via BibTeX.

\end{document}